\documentclass[10pt,letterpaper,aps,prl,twocolumn,superscriptaddress,showpacs]{revtex4-1}
\usepackage{graphicx}
\usepackage{epsfig}
\usepackage{dcolumn}
\usepackage{bm}
\usepackage{times}
\usepackage{mathptmx}
\usepackage{color}
\usepackage{amsmath}
\usepackage{xfrac}
\usepackage{amsfonts}

\begin{document}

\title{A new collective mode in YBCO observed by time-domain reflectometry}

\author{J. P. Hinton}
\affiliation{Materials Science Division, Lawrence Berkeley National Laboratory, Berkeley, California 94720, USA}
\affiliation{Department of Physics, University of California, Berkeley, California 94720, USA}

\author{J. D. Koralek}
\affiliation{Materials Science Division, Lawrence Berkeley National Laboratory, Berkeley, California 94720, USA}

\author{Y. M. Lu}
\affiliation{Department of Physics, University of California, Berkeley, California 94720, USA}

\author{A. Vishwanath}
\affiliation{Materials Science Division, Lawrence Berkeley National Laboratory, Berkeley, California 94720, USA}
\affiliation{Department of Physics, University of California, Berkeley, California 94720, USA}

\author{J. Orenstein}
\affiliation{Materials Science Division, Lawrence Berkeley National Laboratory, Berkeley, California 94720, USA}
\affiliation{Department of Physics, University of California, Berkeley, California 94720, USA}

\author{D. A. Bonn}
\affiliation{Department of Physics and Astronomy, University of British Columbia, Vancouver, Canada.}
\affiliation{Canadian Inst. Adv. Res., Toronto, ON M5G 178, Canada}

\author{W. N. Hardy}
\affiliation{Department of Physics and Astronomy, University of British Columbia, Vancouver, Canada.}
\affiliation{Canadian Inst. Adv. Res., Toronto, ON M5G 178, Canada}

\author{Ruixing Liang}
\affiliation{Department of Physics and Astronomy, University of British Columbia, Vancouver, Canada.}
\affiliation{Canadian Inst. Adv. Res., Toronto, ON M5G 178, Canada}

\begin{abstract}
We report the observation of coherent oscillations associated with charge density wave (CDW) order in the underdoped cuprate superconductor YBa$_{2}$Cu$_{3}$O$_{6+x}$ by time-resolved optical reflectivity. Oscillations with frequency $1.87$ THz onset at approximately $105$~K and $130$~K for dopings of $x = 0.67$ (ortho-VIII) and $x = 0.75$ (ortho-III), respectively. Upon cooling below the superconducting critical temperature ($T_{c}$), the oscillation amplitude is enhanced, the phase shifts by $\pi$, and the frequency softens by $\delta \nu / \nu \approx 7 \% $. A bi-quadratically coupled Landau-Ginzburg model qualitatively describes this behavior as arising from competition between superconducting and CDW orders.
\end{abstract}

\date{\today}

\maketitle

The concept of spontaneous symmetry breaking (SSB) constitutes a paradigm in many branches of physics, most especially high-energy particle phenomenology, and yet has its roots in the study of condensed matter~\cite{NambuPR61,GoldstoneINC61,AndersonPR62}. In metals characterized by strongly-interacting electrons, breaking of U(1) gauge symmetry leads to superconductivity (SC), and breaking of translation and time-reversal symmetry correspond to the formation of charge and spin-density waves. New phenomena may be anticipated in metals in which charge-density waves (CDWs) and SC coexist and interact strongly, such as the coupling of their respective collective modes and possible detection of the SC amplitude mode~\cite{SooryakumarPRL80,LittlewoodPRL81}, the condensed matter analog of the Higgs boson.

Recently, great interest has been generated by the observation, using resonant~\cite{GhiringhelliSCIENCE12,AchkarPRL12,Blanco-CanosaARXIV13} and hard~\cite{ChangNATPHYS12,BlackburnPRL13} X-ray scattering, of a CDW coexisting with SC in underdoped crystals of the prototypical high-$T_{c}$ cuprate superconductor, YBa$_{2}$Cu$_{3}$O$_{6+x}$ (YBCO). Particularly relevant is the finding that the CDW amplitude increases initially as the temperature ($T$) is lowered, but then decreases as $T$ crosses $T_{c}$, indicating a repulsive interaction between SC and CDW phases. These discoveries raise key questions concerning the relation of the newly-found CDW in YBCO to the coupled charge and spin (stripe) order in La$_{2-x}$(Sr, Ba)$_{x}$CuO$_{4}$ and related systems~\cite{TranquadaNATURE95,AbbamonteNATPHYS05}, other phenomena that onset in a similar range of $T>T_c$, such as fluctuating SC~\cite{CorsonNATURE99,XuNATURE00}, anomalies in transient reflectivity, Hall, and Kerr effects~\cite{HeSCIENCE11,LeBoeufNATURE07,XiaPRL08}, and to the opening of the pseudogap itself.

\begin{figure}[ht]
\begin{center}
\includegraphics[width=8.5cm]{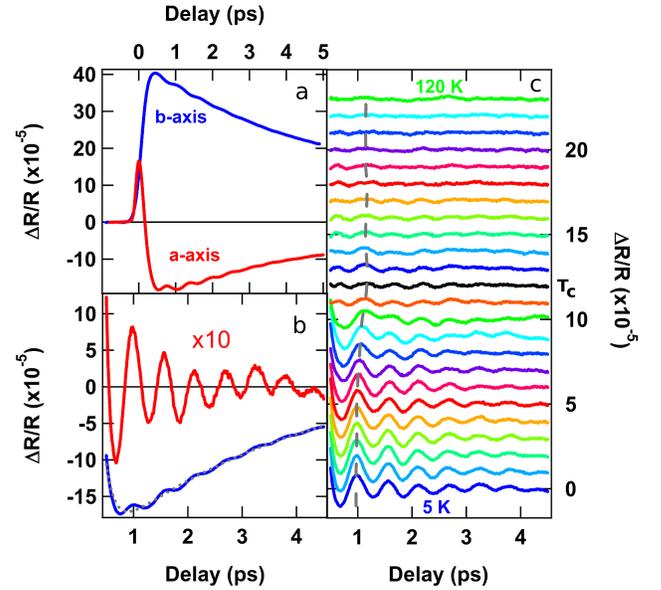}
\end{center}
\caption{\label{fig:1} (a) $\Delta R (t) / R$ for $x = 0.67$ at $15$~K with probe beam polarized parallel to the crystallographic \textit{a} (red) and \textit{b} (blue) axes. Oscillations are present in both polarization channels. (b) \textit{a}-axis $\Delta R (t) / R$ at $5$~K (blue) with the fit to the non-oscillating background (grey dash) and the subtracted oscillating component $\Delta R_{\Omega}$ (red). (c) $T$ dependence of $\Delta R_{\Omega}$, shown in $5$~K steps between $5$ and $120$~K. The first maxima are marked by the grey dash, and the curve nearest $T_{c}$ is plotted in black.}

\end{figure}

In this work, we report the observation by time-domain reflectometry of a new collective mode in YBCO crystals in which the CDW order was previously detected. On the basis of the approximate coincidence of the onset $T$ of the transient reflectivity signal and X-ray scattering, we associate the new mode with the presence of CDW amplitude. From analysis of the time-domain data, we obtain the $T$ dependence of the amplitude, frequency, phase, and damping parameter of the mode. We observe clear anomalies in these parameters as $T$ is lowered in the SC state, providing information concerning the nature of the coupling between the two coexisting forms of order.

Time-resolved measurements of the photoinduced change in optical reflectivity, $\Delta R$, were performed using a mode-locked Ti:Sapphire oscillator generating pulses of 800 nm wavelength light of duration of 60 fs and repetition rate 80 MHz. Measurements were carried out using both the conventional two-beam pump/probe approach and a four-beam transient grating (TGS) configuration~\cite{Goodno,Maznev}. All measurements reported in this Letter were performed using a pump fluence of $1.5 \mu$J$/$cm$^{2}$. Single crystals of YBCO ortho-VIII, with $x = 0.67$, hole doping $p = 0.12$, and $T_{c} = 67$ K, and ortho-III, with $x = 0.75$, $p = 0.13$, and $T_{c} = 75$ K, were studied. By comparing measurements under photoexcitation with thermodynamic and transport measurements, we estimate average laser heating of $5$~K in the vicinity of $T_{c}$.

The three panels of Fig. 1 present results for $\Delta R (t)$ normalized by the equilibrium reflectance $R$ in a crystal of YBCO ortho-VIII, measured using the two-beam configuration.  Fig. 1a illustrates the pronounced anisotropy of the response in the crystallographic \textit{a}-\textit{b} plane, with $\Delta R$ changing sign as the probe polarization is rotated from the \textit{a} to the \textit{b} axis, while the pump polarization is held parallel to the \textit{a} axis (the crystallographic axes were oriented by X-ray Laue diffraction to within $5^{\circ}$). An oscillating reflectivity modulation is clearly present in both polarization channels and is highlighted using the expanded scale of Fig. 1b. In order to quantify the parameters of the oscillation, we subtract a background function of the form $A e^{t/\tau_{1}} + B e^{-t/\tau_{2}}+C$ from each $\Delta R (t) / R$ curve to obtain the oscillating component $\Delta R_{\Omega}$. The resulting difference transient is shown for $T=$5 K in Fig. 1b and for a series of temperatures between $5$ and $120$~K in Fig. 1c.

\begin{figure}[ht]
\begin{center}
\includegraphics[width=8.5cm]{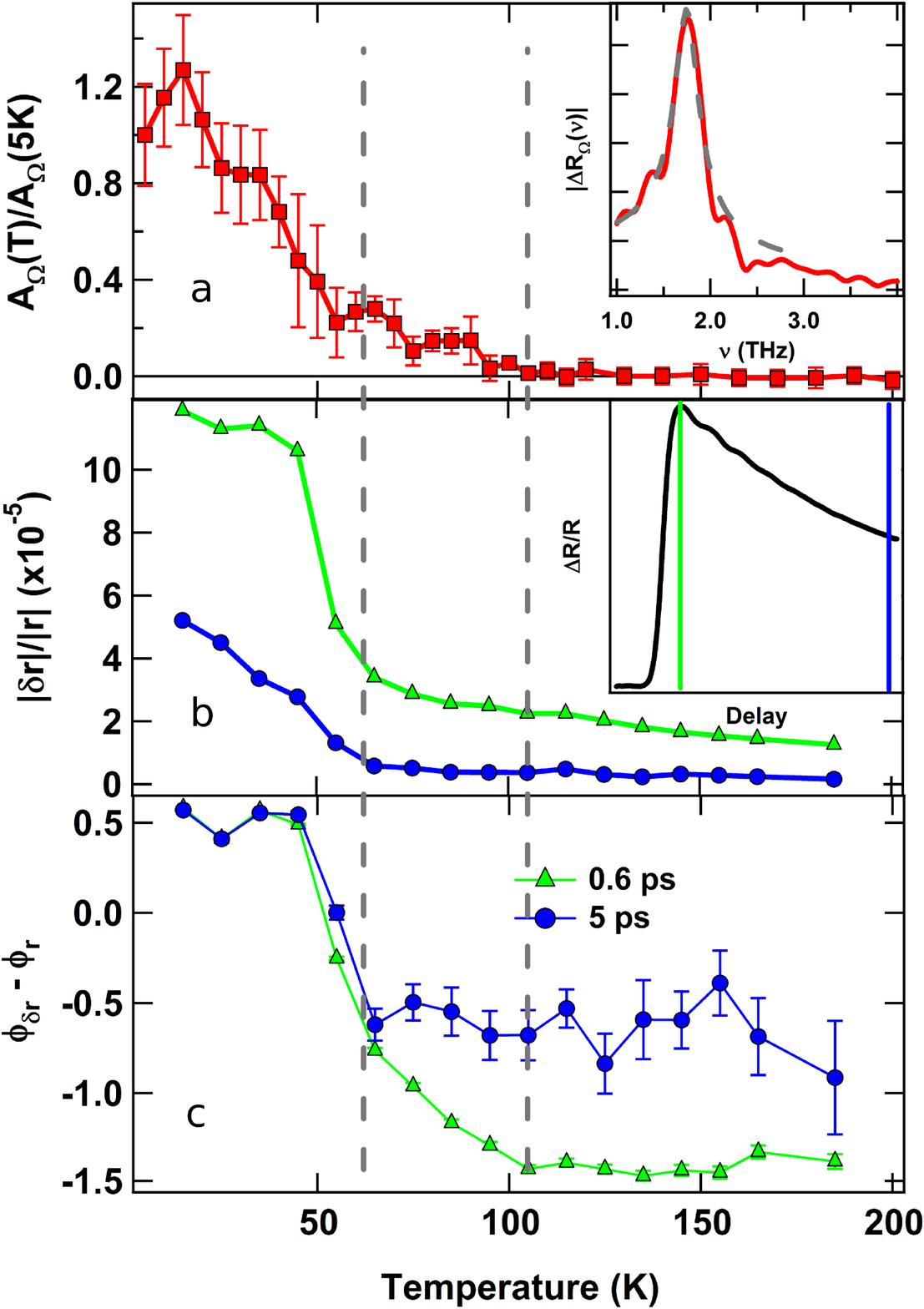}
\end{center}
\caption{\label{fig:2} (a) The $T$ dependence of the oscillation amplitude extracted from the Fourier transforms of the data in Fig. 1c. $\Delta R_{\Omega}(\nu)$ at $5$~K is inset, with the fit in gray dash. The $T$ dependence of (b) the amplitude and (c) phase of the a-axis TGS signal at probe delays of $0.6$ ps and $5$ ps, marked on a representative $\Delta R(t)/R$ curve in the inset to (b). The amplitude increases sharply at $T_{c}$ for all time delays, indicating the onset of the SC response. A phase rotation at $0.6$ ps onsets at $105$~K, while remaining constant in the normal state at $5$ ps. Note the onset of oscillations at $105$~K and the enhancement at $T_{c}$, corresponding to the two onset temperatures observed in the TGS dynamics.}
\end{figure}

Fig. 2a shows the $T$ dependence of the oscillation amplitude, $A_\Omega (T)$ as extracted from the data shown in Fig. 1c. To obtain $A_{\Omega} (T)$ we first Fourier transform $\Delta R_{\Omega}(t)$ to obtain the spectral density function $|\Delta R_{\Omega}(\nu)|^2$ and fit it to a damped harmonic oscillator response function. The spectrum at 5 K and the best fit are shown in the inset to Fig. 2a.  We observe that $A_\Omega (T)$ first appears above the noise level at $T \approx 105$ K and begins to increase much more rapidly upon cooling below $T_{c}$.

Next, we compare $A_\Omega (T)$ with the temperature dependence of the non-oscillatory component of $\Delta R$.  The pump-probe reflectivity dynamics of YBCO are similar to those in other cuprate superconductors~\cite{DemsarPRL99A,KaindlSCIENCE00,GedikSCIENCE03,GedikPRL05,SmallwoodSCIENCE12,CoslovichPRL13,HintonPRL13}; a rapidly decaying transient is observed in the normal state, which persists into the superconducting (SC) state, and a larger and longer-lived response appears below $T_{c}$.  The existence of multiple components below $T_c$ is especially clear in the \textit{a}-axis transient shown in Fig. 1a, where the two components of $\Delta R$ have opposite signs.  The component of $\Delta R$ that appears at or near $T_c$ is associated with the ultrafast photoinduced evaporation of a fraction of the superfluid condensate and its subsequent reformation~\cite{SegrePRL02,KaindlPRB05,KabanovPRL05}. The transient that is observed in the normal state is more complicated, consisting typically of at least two components: a nearly $T$ independent “bolometric” signal associated with photoinduced heating of the electron gas and another component that appears at or below the pseudogap temperature $T^{*}$~\cite{HeSCIENCE11,HintonPRL13}.

We have found that performing time-domain reflectivity in the four-beam TGS mode is very useful in helping to distinguish the several contributions to $\Delta R$ (smooth background and oscillatory) on the basis of their optical phase.  In TGS, the reflected probe pulse is coherently mixed with a local oscillator pulse with an adjustable optical phase, allowing direct measurement of both the amplitude and phase of the photoinduced change in the complex reflection amplitude, $\delta\tilde{r}$~\cite{GedikOPTLETT04}. When applied to YBCO, TGS reveals that, despite the appearance of $\Delta R$ in Fig. 1a, the non-oscillatory components of the transient reflectivity $\delta\tilde {r}_{a}$ and $\delta\tilde{r}_{b}$ are identical, with the apparent anisotropy coming entirely from a phase difference between $\tilde{r}_a$ and $\tilde{r}_b$. However, analysis of the TGS data indicates that the oscillatory components do display some anisotropy; the ratio of the oscillatory to background components of $\delta\tilde{r}$ is $35 \pm 18 \%$ larger along \textit{a} than along \textit{b}. Finally, TGS shows that the oscillations are not a modulation of the background $\delta\tilde{r}$, since the two components of the signal have distinct optical phases, by $0.3 \pi$.

In Figs. 2b and 2c we plot the amplitude and optical phase (relative to $\tilde{r}_a$), respectively, of $\delta\tilde{r}_a$ as a function of $T$, for two representative time delays. The amplitude, $\lvert\delta\tilde{r}\rvert$, increases sharply at $T_c$ for both delays, in a manner that appears to be correlated with the amplitude of the oscillation. Above $T_c$, $\lvert\delta\tilde{r}\rvert$ displays a weak and featureless $T$ dependence, persisting above $200$~K. Despite the smooth $T$-dependence of the $\delta\tilde{r}$ amplitude, the phase (Fig. 2c) shows structure upon cooling below $105$~K, the same $T$ at which the reflectivity oscillations become observable. At this point we cannot tell whether this feature in the phase of $\delta \tilde{r}$ reflects the onset of CDW order or phase-fluctuating superconductivity.

\begin{figure}[ht]
\includegraphics[width=8.5cm]{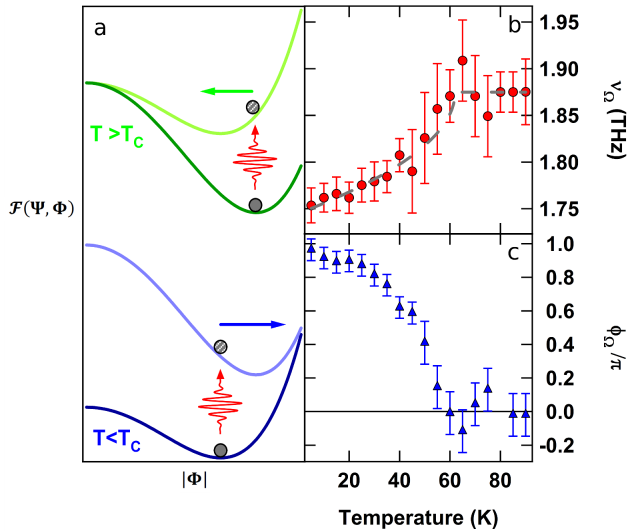}
\caption{\label{fig:3} (a) Schematic representation of displacive excitation mechanism described in the text. Below $T_{c}$, pump excitation enhances CDW order via partial evaporation of the SC condensate, as represented schematically in the lower half of (a). Above $T_{c}$, the dominant process is direct photo-suppression of CDW order, leading to a reversal in the initial displacement of the free energy minimum. (b) and (c) give the $T$ dependence of the frequency and phase of the CDW oscillations,respectively. The grey dash in (b) gives the fit to $\nu_{\Omega}(T)$ obtained from coupled Ginzburg-Landau model.}
\end{figure}

The most striking aspect of data plotted in Fig. 2 is the enhancement $A_{\Omega}(T)$ correlated with the onset of superconductivity. The result is surprising given the direct observation by X-ray scattering of the reduction of the CDW amplitude with the appearance of superconductivity in the same material ~\cite{ChangNATPHYS12}. Since we associate $\Delta R_{\Omega}$ with the onset of CDW order, it would be natural to expect a suppression of the collective mode due to competition with superconductivity.  Below we show how this seemingly anomalous behavior can emerge as a consequence of SC-CDW coupling.

A simple description of coupled order parameters is given by the Ginzburg-Landau free energy,
\begin{equation}
\mathcal{F}(\Phi, \Psi) = -a \lvert \Phi \rvert ^{2} + \frac{b}{2}\lvert \Phi \rvert^{4} - \alpha \lvert \Psi \rvert^{2} + \frac{\beta}{2} \lvert \Psi \rvert^{4} + \lambda \lvert \Phi \rvert^{2} \lvert \Psi \rvert^{2}
\end{equation}
where $\Phi$ and $\Psi$ are the CDW and SC order parameters, respectively, and $\lambda$ is a constant representing the strength of coupling between the two. For the case of coexisting orders with a repulsive interaction, the parameters $a$, $\alpha$, and $\lambda$ are all positive.  Minimizing the free energy with respect to $\lvert \Phi \rvert$ yields for the equilibrium value of the CDW order parameter,
\begin{equation}
\lvert \Phi_{eq} \rvert ^{2} = \frac{a-\lambda \lvert \Psi \rvert^{2}}{b},
\end{equation}
 showing that for a repulsive interaction between the two orders, the CDW amplitude is suppressed in the presence of SC order.

Fig. 3a shows how the repulsive interaction between the two order parameters provides a mechanism for photoexcitation of the CDW-related mode in the SC state.  The lower part of this figure ($T<T_{c}$) is a plot of the free energy as a function of the CDW order parameter for two values of the SC amplitude.  The curve with the lowest free energy corresponds to the equilibrium value of $\Psi$ while the curve above it is calculated using a smaller value of $\Psi$.  As expected from the repulsive interaction between the two orders, the CDW amplitude is larger in the state with weaker SC.  As illustrated in Fig. 3a by the directed laser pulse, the effect of photoexcitation is to impulsively reduce the SC order, inducing a transition from the lower to upper free energy curves.  The sudden shift in the value of $\lvert \Phi \rvert$ that minimizes the free energy drives the oscillation of the CDW amplitude.

The initial amplitude of the ensuing oscillation is the shift in the minimum of $\Phi$ that accompanies photoexcitation, which to first order in $\lambda$ is equal to,
\begin{equation}
\delta \lvert \Phi \rvert = \lvert \Phi_0 \rvert \frac{\lambda}{2 a}\delta\lvert \Psi \rvert^{2}\propto \lvert\Psi\rvert^2,
\end{equation}
where $\Phi_0$ is the equilibrium value CDW order in the absence of coupling to SC. The proportionality to $\lvert\Psi\rvert^2$ on the right-hand side of Eq. 3 follows from the previous demonstration~\cite{KaindlPRB05} that, for all temperatures below $T_c$, the photoinduced decrease in SC order for a fixed laser fluence is proportional to $\lvert\Psi\rvert^2$ itself. This simple calculation reproduces the counter-intuitive result that the amplitude of the CDW oscillations is proportional to $\lvert\Psi\rvert^2$, and therefore is enhanced by superconductivity, even as $\lvert \Phi_{eq} \rvert$ is suppressed.

The $T$ dependence of the oscillation frequency $\nu_{\Omega}$ obtained from the fits to $\Delta R_{\Omega}(\nu)$ is plotted in Fig. 3a. The frequency in the normal state, 1.87 THz (62 cm$^{-1}$) is well below that of the lowest optic phonon in the YBCO system, which is a Ba-O mode whose frequency is $\approx$120 cm$^{-1}$, as measured both by both Raman~\cite{Friedl} and time-domain reflection spectroscopies~\cite{Albrecht,Mazin,Misochko}. As Fig. 3a shows,$\nu_{\Omega}$ begins to decrease with the appearance of SC order, accompanied by a decrease in the damping time from $1.9 \pm 0.5$ ps to $1.3 \pm 0.2$ ps, further evidence for a picture of coupled order parameters. Softening of the restoring force of the oscillation is expected from the curvature of $\mathcal{F}$ evaluated at equilibrium value of CDW order,
\begin{equation}
\frac{\partial^2\mathcal{F}}{\partial\lvert\Phi\rvert^2}\bigg|_{eq} = 4(a-\lambda \lvert \Psi \rvert^{2}).
\end{equation}
The dashed curve through the data points in Fig. 3b is a fit to Eq. 4 assuming a temperature dependent SC order parameter $\lvert \Psi \rvert^{2} (T) \propto (1-T^{2}/T_{c}^{2})^{1/2}$.

Another clear, and yet unexpected, feature of the $T$ dependence of the collective mode parameters is the shift in the phase $\phi_{\Omega}$ of the oscillations upon entering the SC state, as shown in Fig. 3c (note that this phase is different from the optical phase of the reflectivity amplitude discussed above).  The phase shift is apparent from the time-domain oscillations plotted in Fig. 1c, where the dotted vertical line traces out the delay time of the first crest of the oscillation as a function of $T$. As indicated in Fig. 3c, total phase difference from $T>T_c$ to low temperature is quite close to $\pi$. To explain this effect, we consider the excitation mechanism of the CDW oscillation in the normal state.  In the absence of a competing SC state, pumping is expected to weaken the CDW order~\cite{TorchinskyNATMAT13,StojchevskaPRB11,DemsarPRL99B}.  This corresponds to a decrease in the quasi-equilibrium value of $\lvert \Phi \rvert$, as shown in the upper part of Fig. 3a, where we plot $\mathcal{F}$ as a function of $\lvert\Phi\rvert$ in equilibrium (lower curve) and after photoexcitation (upper curve).  As the horizontal arrows indicate, the sign of the initial displacement above $T_{c}$ is opposite to that predicted in the SC state, in agreement with our observation of a $\pi$ phase shift.

\begin{figure}[ht]
\begin{center}
\includegraphics[width=8.5cm]{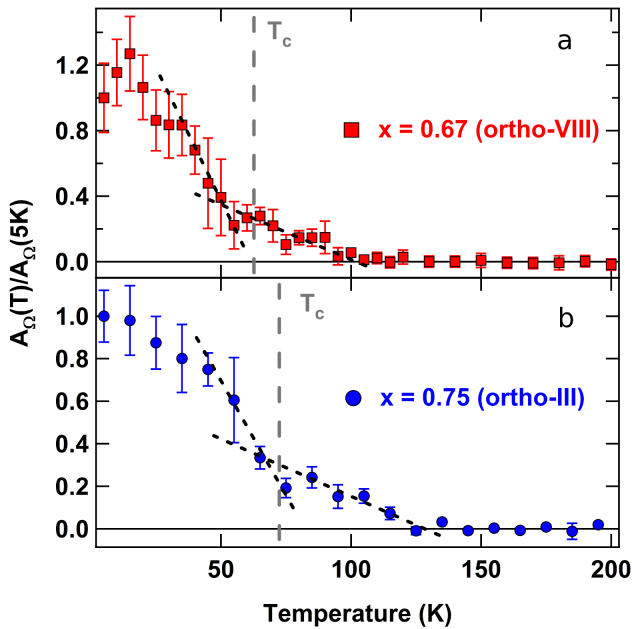}
\end{center}
\caption{\label{fig:4}The $T$ dependence of the CDW oscillations for (a) $x = 0.67$ (ortho-VIII) and (b) $x = 0.75$ (ortho-III). The black dashed lines are guides to the eye. Oscillations initially onset at approximately $105$~K in ortho-VIII and $130$~K in ortho-III, with both dopings displaying a pronounced enhancement in oscillation amplitude at $T_{c}$.}
\end{figure}

In addition to the measurements reported above, we have observed and characterized the collective mode oscillations in YBCO ortho-III, in which CDW ordering has also been observed.  In Fig. 4 we compare the amplitude of the oscillations in the two compounds that we have studied thus far.  Overall, the behavior of $A_{\Omega}(T)$ is qualitatively the same - onset in the normal state and rapid increase for $T<T_c$ - although in YBCO ortho-III the oscillations first appear at a higher $T \approx 130$ K.  In addition, in the ortho-III compound we have found the same shift in phase and reduction in frequency upon entering the superconducting state as shown for the ortho-VIII sample in Fig. 3.

By way of conclusion, we address some of the questions raised by our observations and their implications for future research. The first such question is whether the mode we have observed is, in fact, the amplitude mode of the CDW. On the basis of the measured frequency of 60 cm$^{-1}$, we can rule out phonon modes of the structure above the CDW ordering temperature, $T_{CDW}$. The fact that the new mode appears only at $T<T_{CDW}$ strongly suggests that it is correlated with symmetry-breaking in a low $T$ phase.  However, the parameters of the oscillation do not show some of the signatures expected for an amplitude mode.  At the level of mean-field theory, the frequency of the amplitude mode should go to zero as $T$ approaches the CDW ordering temperature, $T_{CDW}$.  Although in real CDW systems the frequency does not shift all the way to zero, the mode is found to soften considerably and become overdamped as $T\rightarrow T_{CDW}$~\cite{DemsarPRL99B,TravagliniSSC83}. By contrast, the frequency and damping parameters that we observe remain constant even as the amplitude of the mode vanishes with increasing $T$. This apparent discrepancy raises the possibility that the new mode originates as an acoustic phonon whose wavevector matches that of the CDW.  Once the CDW distortion takes place this mode will be shifted to zero wavevector and become observable by optical probes~\cite{KenjiPRB98}.

A second, and closely related question, is the relation of the new mode in YBCO to low-frequency collective modes observed in the La$_{2-x}($Sr$_{x}$,Ba$_x$)CuO$_{4}$ family of cuprates. In particular, the mode frequency ($\nu\approx 2.0$ THz) and $T$-dependent amplitude recently reported~\cite{TorchinskyNATMAT13} in thin films of La$_{2-x}$Sr$_{x}$CuO$_{4}$ (LSCO) are strikingly similar to what we observe in the YBCO crystals, although the damping time is roughly five times shorter.  The similarity is surprising in that scattering probes suggest very different CDW structures in the two compounds: stripe-like coupled spin and charge fluctuations in LSCO~\cite{TranquadaNATURE95,AbbamonteNATPHYS05}, as opposed to charge density waves uncorrelated with spin fluctuations, perhaps with a 2D checkerboard structure, in YBCO~\cite{GhiringhelliSCIENCE12}. However, the similarity might be explained if the collective modes in both systems result from mixing down of acoustic modes to zero wavevector, rather than CDW amplitude modes. We believe that, regardless of what ultimately proves to be their origin, the careful study of these new collective modes as a function of temperature, doping, and magnetic field will contribute greatly to our understanding of the role of CDW and SC coupling in the cuprate family of superconductors.

The work in Berkeley was supported by the Director, Office of Science, Office of Basic Energy Sciences, Materials Sciences and Engineering Division, of the U.S. Department of Energy under Contract No. DE-AC02-05CH11231.

\end{document}